\documentclass[usenatbib,useAMS,a4paper]{mn2e}
\usepackage{longtable}
\usepackage{graphicx}
\usepackage{txfonts}
\usepackage{natbib}
\usepackage{times}

\newcommand\beq{\begin{equation}}
\newcommand\eeq{\end{equation}}

\newenvironment{acknowledgements}{\section*{Acknowledgements}}{}

\def\institute{
Department of Physics and Astronomy, McMaster University, 1280 Main Street
West, Hamilton, Ontario L8S 4M1, Canada
}

\author[E. Glebbeek et al.]{Evert Glebbeek, Alison Sills and Nathan Leigh\\\institute}

\title[Multiple populations and blue stragglers]{Multiple stellar populations and their influence on blue stragglers}

\def\abstractx{
It has become clear in recent years that globular clusters are not
simple stellar populations, but may host chemically distinct
sub-populations, typically with an enhanced helium abundance. These
helium-rich populations can make up a substantial fraction of all cluster
stars.

One of the proposed formation channels for blue straggler stars is the
physical collision and merger of two stars. In the context of multiple
populations, collisions between stars with different helium abundances
should occur and contribute to the observed blue straggler population.
This will affect the predicted blue straggler colour and luminosity
function.

We quantify this effect by calculating models of mergers resulting from
collisions between stars with different helium abundances and using these
models to model a merger population. We then compare these results to
four observed clusters, NGC 1851, NGC 2808, NGC 5634 and NGC 6093.

As in previous studies our models deviate from the
observations, particularly in the colour distributions. However,
our results are consistent with observations of multiple
populations in these clusters. In NGC 2808, our best fitting models include
normal and helium enhanced populations, in agreement with helium
enhancement inferred in this cluster. The other three clusters show better
agreement with models that do not include helium enhancement. We discuss
future prospects to improve the modelling of blue straggler populations and
the role that the models we present here can play in such a study.
}
\def\keywordsx{
stars: blue stragglers, evolution, luminosity function -- globular
clusters: general
}

\begin{document}
\bibliographystyle{mn2e}

\maketitle

\begin{abstract}
\abstractx
\end{abstract}


\section{Introduction}

Blue stragglers are stars that appear on the extension of the main sequence
in both open and globular clusters and have been studied extensively since
their original discovery by \citet{Sandage1953}. Their existence cannot be
explained by conventional stellar evolution models.
Two formation channels have been discussed extensively in the literature,
one using stellar collisions
\citep{Lombardi1995,Sillsetal1997,Sills2005,GlebbeekPols2008} and one using
binary evolution \citep{ChenHan2004,ChenHan2008,ChenHan2009}.  So far
observations do not clearly indicate which of these formation mechanisms is
the dominant one, or even if there is a dominant formation mechanism
\citep{KniggeLeighSills2009}.
Neither formation mechanism has been able to completely account for the
properties of the observed blue straggler populations, in particular the
colour and magnitude distributions.

Star to star abundance variations in globular clusters, originally known as
``globular cluster abundance anomalies'', have been known for a long time
(see \citealt{Gratton2004} for a recent review).
The observed abundance patterns indicate that at least some of the globular
cluster stars (up to 50\% of the present day population) contain material
that has undergone partial hydrogen burning at high temperatures. The most
famous example is the so-called oxygen-sodium anti-correlation.
To date, abundance variations have been observed in all clusters where the
observational data has sufficient resolution to detect them. Despite work
by several groups, the origin of these abundance variations is still
unclear. The leading models
propose that material has been processed by AGB stars
\citep{Venturaetal2001,Venturaetal2002,Venturaetal2005}
or rapidly rotating massive stars \citep{Decressinetal2007} before being
returned to the interstellar medium to form new stars. An alternative
scenario involves non-conservative mass transfer in a binary
system \citep{deMink2009}.
Since the main product of hydrogen burning is helium, this polluted
material is also expected to be relatively helium rich and we may therefore
expect to find a helium enhanced sub-population among cluster stars.

The first evidence for the existence of multiple populations in gobular
clusters comes from the discovery of a split giant branch in $\omega$ Cen
\citep{Pancino2000}. It was later shown that the sub giant branch
\citep{Ferraro2004} and the main sequence \citep{Bedin2004} are likewise split
in multiple sequences, which is interpreted as a spread
in helium abundance \citep{Piotto2005}.
\citet{dAntona2005} predicted the existence of multiple
populations in NGC 2808 based on the observed width of the
main sequence band. Detailed observations by \citet{Piotto2007} found that
the main sequence of NGC 2808 indeed splits into three distinct sequences.
\citet{Milone2008} reported the existence of a double sub-giant branch in
NGC 1851.

If collisions contribute to blue straggler formation and the different
sub-populations have a comparable number of stars, then collisions
between stars from a helium-rich and a helium-normal population must
contribute to the observed blue straggler population.
A blue straggler that is formed from a collision between a helium normal
and a helium-rich star will have a higher helium abundance in its interior
than a blue straggler formed by a collision between equivalent stars with
the same helium abundance. This has the effect of making the star
brighter, even if the material is not mixed efficiently
\citep{GlebbeekPH2008}.
Therefore, collisions involving stars from different populations should be
taken into consideration when constructing the blue straggler colour
and luminosity functions in clusters with multiple populations.

Here we explore the results of such a calculation.

\section{Computational set-up}
We use the STARS code (\S \ref{sec:stars}) to calculate the evolution
tracks of both the progenitor stars and the collision products. The
structure of the collision products themselves is calculated using the MMAS
code (\S \ref{sec:mmas}).

The parameter space for our models consists of the masses of the two
colliding stars, $M_1$ and $M_2$, the time of collision $t_\mathrm{coll}$
and the helium abundance of the colliding stars, $Y_1$ and $Y_2$. In
principle the impact parameter for the collision spans another dimension in
parameter space, but we will limit ourselves to non-rotating collision models
here.
In essence, we assume that the collision product has an efficient
way to lose excess angular momentum (\emph{e.g.} by the action of a magnetic
field) and that rotational mixing is unimportant.
With this assumption, the effect of a non-zero impact parameter on
the structure of the collision product is small \citep{Sillsetal1997}
and can be ignored.

We calculated a grid of models that maps out the relevant portion of
parameter space for blue straggler formation in globular clusters.
Our computational grid covers a mass range of $0.4\ldots
0.8\mathrm{M}_\odot$ for both stars. We will refer to the most massive star
as the primary and the least massive one as the secondary.
To account for the evolution of the parent stars as well as the evolution of
the collision product since time of collision the collision time
is varied between $6\,000 \,\mathrm{Myr}$ and $12\,000 \,\mathrm{Myr}$ in
steps of $1\,000 \,\mathrm{Myr}$.
The heavy element content $Z$ is set to $Z = 0.001$ for most of
our models and abundances are scaled to solar values. The initial (ZAMS)
helium abundance in the stars is set to $Y = Y_0 + 2 Z$, where $Y_0$ is
$0.24, 0.32$ or $0.40$. This parameter range was chosen to cover both
normal helium abundances ($Y_0 = 0.24$) and the most extreme helium
abundance proposed in the literature ($Y_0 = 0.40$, in NGC 2808,
\citealt{dAntona2005,Piotto2007}).
The initial hydrogen abundance $X = 1 - Y - Z$ and the primary and
secondary are allowed to have a different $Y_0$.  The range of parameters
in our grid is summarised in Table \ref{tab:grid}.

\begin{table}
\caption{Parameters for the different model sets}
\begin{tabular}{llll}
\hline
Model set & $Y_0$ & $t_\mathrm{coll}$ & $M_1, M_2$ \\
          &       & {Myr} & {$\mathrm{M}_\odot$}\\
\hline
\hline
A     & $0.24$             & $6\,000$ -- $12\,000$ & $0.4$, $0.6$, $0.8$\\
B     & $0.32$             & $6\,000$ -- $12\,000$ & $0.4$, $0.6$, $0.8$\\
C     & $0.40$             & $6\,000$ -- $12\,000$ & $0.4$, $0.6$, $0.8$\\
D     & $0.24;0.32$        & $6\,000$ -- $12\,000$ & $0.4$, $0.6$, $0.8$\\
E     & $0.24;0.32;0.40$   & $6\,000$ -- $12\,000$ & $0.4$, $0.6$, $0.8$\\
\hline
\end{tabular}
\label{tab:grid}
\end{table}

The model sets A, B and C all involve just collisions between stars of the
same helium content. Model set D also includes collisions between stars with
$Y_0 = 0.24$ and $Y_0 = 0.32$ and is a superset of models A and B.
Similarly model set E is a superset of model sets A, B and C that also
contains the cross-collisions involving stars with $Y_0 = 0.40$.

Model sets A, B and D have been calculated for $Z = 0.0003$ as
well as $Z = 0.001$.

\subsection{MMAS}\label{sec:mmas}
The structure of our collision products was calculated using the Make Me A
Star (MMAS) code by \citet{Lombardietal2002}. MMAS approximates the
structure of a collision product using an algorithm known as entropy
sorting.

The idea behind this algorithm is that the quantity $A = p/\rho^{5/3}$,
which is related to the thermodynamic entropy, increases (nearly)
monotonically inside stars from centre to surface. In the absence of strong
shocks, as is the case for low-velocity collisions, $A$ is conserved in the
fluid elements from both colliding stars. The structure of the collision
product can then be approximated by sorting the mass shells of both parent
stars in order of increasing $A$.
As a result of stellar evolution $A$ decreases in the core, so that the
core of the collision product is most likely to resemble the core of the
most evolved parent star \citep{Sillsetal1997,GlebbeekPols2008}.

\subsection{The STARS code}\label{sec:stars}
Our evolutionary models are calculated using a version of Eggleton's
stellar evolution code \citep{Eggleton1971, Polsetal1995}, hereafter STARS.
The STARS code solves the equations of stellar structure and the nuclear
energy generation rate simultaneously on an adaptive non-Lagrangian
non-Eulerian (``Eggletonian'') grid \citep{Stancliffe2006}.
Since \citet{GlebbeekPH2008} we have update the nuclear reaction rates
to the recommended values from \citep{NACRE}, with the exception of the
$^{14}\mathrm{N} (\mathrm{p}, \gamma)^{15}\mathrm{O}$ reaction, for which
we use the recommended rate from \citet{Herwigetal2006} and
\citet{Formicola2004}.
We use the opacity tables of \citet{EldridgeTout2004}, which combine the
OPAL opacities from \citet{IglesiasRogers1996}, the low temperature
molecular opacities from \citet{AlexanderFerguson1994}, electron scattering
opacities from \citet{BuchlerYueh1976} and the conductive opacities from
\citet{Iben1975}.
The assumed heavy-element composition is scaled to the solar mixture
of \citep{AndersGrevesse1989}.
Chemical mixing due to convection
\citep{BohmVitense1958,Eggleton1972} and
thermohaline mixing \citep{Kippenhahnetal1980, Stancliffe2007} is taken
into account. 
All models are computed with a mixing-length ratio $l/H_P=2.0$.  As in
\citet{GlebbeekPols2008}, we have neglected convective overshooting in all
models described here.
Its effect on the evolution of stars in the mass range of our collision
products ($0.8 \mathrm{M}_\odot$ -- $1.6 \mathrm{M}_\odot$) is negligible
\citep{Pols1998}.

On the giant branch and the early AGB we adopt the Reimers-like mass loss
prescription from \citet{SchroederCuntz2005,SchroederCuntz2007},
\beq\label{eqn:scmassloss}
\dot M = \eta \frac{L R}{M}
\left(\frac{T_\mathrm{eff}}{4000 \mathrm{K}}\right)^{3.5}
\left(1 + \frac{g_\odot}{4300 g}\right).
\eeq

Because we cannot calculate through the core helium flash, we stop the code
at the onset of the flash and construct a ``zero-age horizontal branch''
model with the correct total mass, core mass and composition.
This construction is performed by pseudo-evolving a low mass core
burning helium star.
Composition changes due to helium burning are disabled during this process,
but hydrogen is allowed to burn normally so that the helium core can grow
in mass. 
Material of the appropriate envelope composition is accreted on the star
until its mass matches that of the last pre-flash model.
The helium core is then allowed to grow to the desired mass due to hydrogen
shell burning until the helium core mass too matches the mass in our
pre-flash model. This procedure gives us a core helium burning star of the
correct total mass, core mass and composition, but we ignore any mass loss
or mixing that may arise as a result of the helium flash itself. This is
not an unreasonable assumption \citep{HarmSchwarzschild1966,Dearborn2006}
although some authors have found mixing during calculations of the helium
flash at very low metallicity \citep{Fujimoto1990,Campbell2008}.
Recent 3D hydrodynamical calculations of the helium flash by
\citet{Mocaketal2009} are consistent with this picture, but suggest that
the structure of the star on the approach of the helium flash may be
different than earlier calculations predict.

We make no attempt to follow thermal pulses during the AGB phase and most
of our models terminate at the beginning of the tp-AGB.

\section{The structure and evolution of helium enhanced stars}
\label{sec:he_enhanced_stars}
Before we consider the evolution of mergers resulting from collisions
involving helium enhanced stars, we review the properties of helium-rich
stars and compare them to helium normal ($Y_0 = 0.24$) stars.
These results are summarised in Tables \ref{tab:normal_properties_y24} --
\ref{tab:normal_properties_y40}.

\begin{table}
\caption{Properties of helium normal ($Y_0 = 0.24$) stars. Masses are in
solar units, lifetimes in Myr.}
\begin{center}
\begin{tabular}{lrrrrrr}
\hline
$M$ & $\tau_\mathrm{MS}$ & $\tau_\mathrm{RGB}$ & $\tau_\mathrm{HB}$ &
$\tau_\mathrm{AGB}$ & $M_\mathrm{c, tGB}$ & $M_\mathrm{c, eAGB}$\\
\hline\hline
 0.60 &  37601 &  2751.1 &  185.0 &  7.70 &  0.47 &  0.38\\
 0.71 &  18999 &  2264.3 &  109.4 &  5.00 &  0.49 &  0.42\\
 0.79 &  11934 &  1976.6 &  107.0 &  4.30 &  0.49 &  0.43\\
 0.89 &  8414 &  722.7 &  107.2 &  4.55 &  0.49 &  0.43\\
 1.00 &  5042 &  1019.7 &  106.8 &  4.20 &  0.49 &  0.44\\
 1.10 &  3610 &  782.5 &  104.5 &  3.56 &  0.49 &  0.45\\
 1.20 &  2920 &  399.9 &  103.8 &  3.42 &  0.48 &  0.45\\
 1.30 &  2303 &  210.0 &  103.3 &  3.64 &  0.48 &  0.46\\
 1.40 &  1831 &  122.0 &  102.4 &  3.10 &  0.48 &  0.46\\
 1.50 &  1516 &  72.8 &  105.4 &  3.36 &  0.47 &  0.46\\
 1.60 &  1277 &  45.6 &  107.5 &  2.87 &  0.47 &  0.47\\
\hline
\end{tabular}
\end{center}
\label{tab:normal_properties_y24}
\end{table}

\begin{table}
\caption{As Table \ref{tab:normal_properties_y24} for $Y_0 = 0.32$.}
\begin{center}
\begin{tabular}{lrrrrrr}
\hline
$M$ & $\tau_\mathrm{MS}$ & $\tau_\mathrm{RGB}$ & $\tau_\mathrm{HB}$ & $\tau_\mathrm{AGB}$ & $M_\mathrm{c, tGB}$ & $M_\mathrm{c, eAGB}$\\
\hline\hline
 0.60 &  21986 &  2389.3 &  184.8 &  8.00 &  0.47 &  0.38\\
 0.71 &  11072 &  1745.1 &  110.7 &  4.20 &  0.48 &  0.43\\
 0.79 &  7080 &  1338.5 &  106.7 &  3.57 &  0.47 &  0.45\\
 0.89 &  4711 &  871.7 &  105.6 &  3.11 &  0.47 &  0.46\\
 1.00 &  3066 &  711.4 &  103.9 &  2.89 &  0.47 &  0.48\\
 1.10 &  2345 &  368.7 &  103.4 &  3.02 &  0.47 &  0.48\\
 1.20 &  1810 &  212.0 &  102.1 &  2.56 &  0.47 &  0.49\\
 1.30 &  1443 &  125.4 &  102.6 &  2.68 &  0.46 &  0.49\\
 1.40 &  1195 &  65.1 &  104.8 &  2.26 &  0.45 &  0.51\\
 1.50 &  981 &  48.1 &  114.9 &  2.23 &  0.43 &  0.51\\
 1.60 &  838 &  29.7 &  128.2 &  0.85 &  0.39 &  0.51\\
\hline
\end{tabular}
\end{center}
\end{table}

\begin{table}
\caption{As Table \ref{tab:normal_properties_y24} for $Y_0 = 0.40$.}
\begin{center}
\begin{tabular}{lrrrrrr}
\hline
$M$ & $\tau_\mathrm{MS}$ & $\tau_\mathrm{RGB}$ & $\tau_\mathrm{HB}$ & $\tau_\mathrm{AGB}$ & $M_\mathrm{c, tGB}$ & $M_\mathrm{c, eAGB}$\\
\hline\hline
 0.60 &  12336 &  1726.1 &  168.5 &  7.60 &  0.46 &  0.39\\
 0.71 &  6438 &  1009.2 &  105.8 &  3.43 &  0.46 &  0.46\\
 0.79 &  4220 &  714.8 &  100.9 &  2.68 &  0.46 &  0.49\\
 0.89 &  2644 &  657.0 &  103.0 &  2.26 &  0.45 &  0.51\\
 1.00 &  1938 &  294.9 &  100.1 &  1.99 &  0.45 &  0.53\\
 1.10 &  1485 &  156.8 &  98.3 &  1.82 &  0.45 &  0.55\\
 1.20 &  1174 &  84.5 &  99.4 &  1.71 &  0.44 &  0.56\\
 1.30 &  940 &  53.6 &  103.2 &  1.58 &  0.42 &  0.57\\
 1.40 &  776 &  33.7 &  102.5 &  0.73 &  0.39 &  0.58\\
 1.50 &  646 &  23.9 &  119.6 &  0.63 &  0.35 &  0.44\\
 1.60 &  547 &  17.8 &  111.6 &  1.28 &  0.34 &  0.61\\
\hline
\end{tabular}
\end{center}
\label{tab:normal_properties_y40}
\end{table}

These tables give the lifetimes on the main sequence ($\tau_\mathrm{MS}$),
the red giant branch ($\tau_\mathrm{RGB}$), the horizontal branch
($\tau_\mathrm{HB}$) and the AGB ($\tau_\mathrm{AGB}$, not including the
tp-AGB). We also give the core mass at the tip of the giant branch
($M_\mathrm{c, tGB}$) and at the beginning of the AGB ($M_\mathrm{c,
eAGB}$).

\begin{figure}
\includegraphics[width=0.5\textwidth]{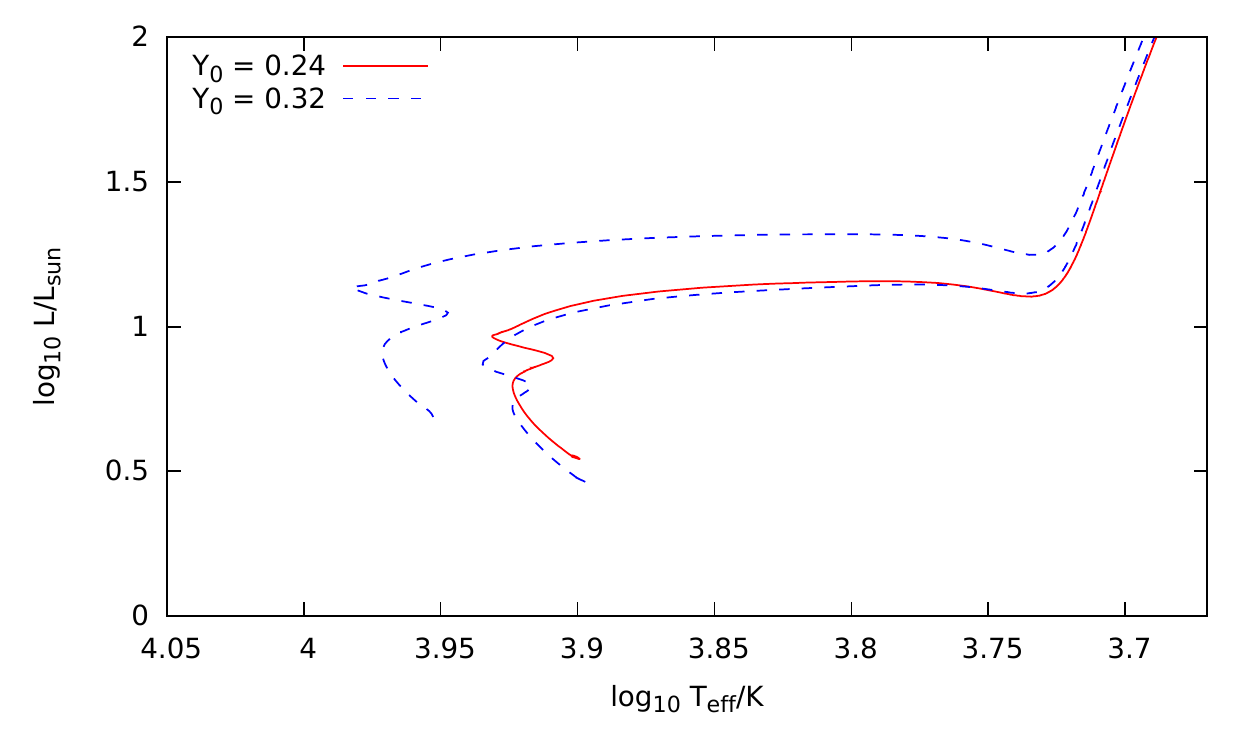}
\caption{
Evolution tracks for a $1.2 \mathrm{M}_\odot$ helium-normal star (solid
track) and a $1.2 \mathrm{M}_\odot$ (upper dashed track) and a $1.05
\mathrm{M}_\odot$ (lower dashed track) $Y_0 = 0.32$ star.
}\label{fig:compare_helium_tracks_y32}
\end{figure}

\begin{figure}
\includegraphics[width=0.5\textwidth]{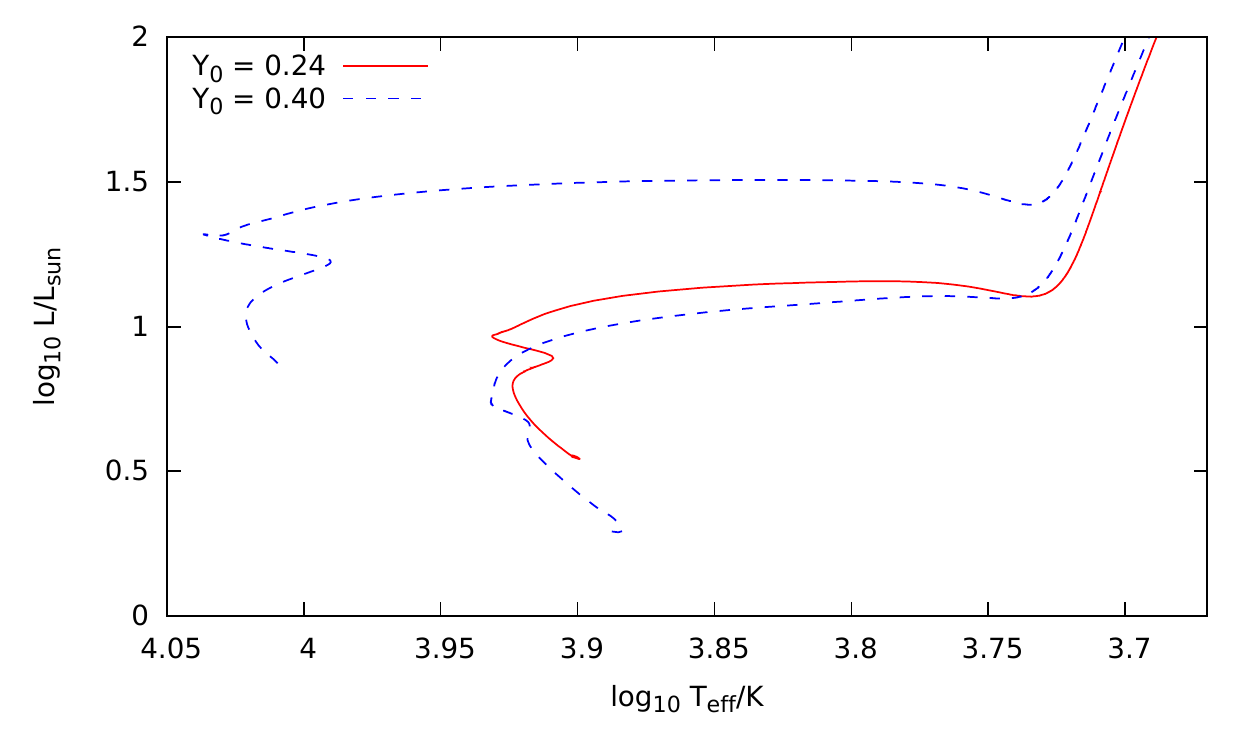}
\caption{
As figure (\ref{fig:compare_helium_tracks_y32}) for $Y_0 = 0.4$. The lower
dashed track is for a $0.9 \mathrm{M}_\odot$ star.
}\label{fig:compare_helium_tracks_y40}
\end{figure}

Perhaps the best-known property of helium-rich stars is that they are
hotter and brighter than helium-normal stars of the same mass (Figures
\ref{fig:compare_helium_tracks_y32} and \ref{fig:compare_helium_tracks_y40}).
In this sense helium-rich stars can also masquerade as more metal-poor
stars, which are also hotter and brighter than metal-rich stars. This has
been used to explain the existence of a blue metal \emph{rich} sequence in
$\omega$ Cen \citep{Piotto2005} and more recently to explain the split subgiant
branch in NGC 1851 \citep{Hanetal2009}.

A direct consequence of their higher luminosity is that a helium-rich star of
a given mass has a considerably shorter main sequence lifetime than a
helium normal star of the same mass. A helium-rich star with the same
lifetime as a helium normal star will be less massive.
The evolution tracks of such ``coeval'' stars are also much closer in terms
of colour and luminosity (Figures \ref{fig:compare_helium_tracks_y32} and
\ref{fig:compare_helium_tracks_y40}).
In other words, although it is true that helium-rich stars are brighter and
bluer than normal stars for the same mass, this effect is much less
significant when comparing populations of stars with the same age.
This is the reason that a helium-rich population
in a cluster shows up as a broadening (or splitting) of the main sequence
rather than as an extension of the main sequence.

Helium-rich stars mimic helium normal stars of higher mass in other ways.
For $Y_0 = 0.24$, a convective core appears on the main sequence above $M =
1.1 \mathrm{M}_\odot$, while for $Y_0 = 0.32$ an $M = 1 \mathrm{M_\odot}$
solar mass star already has a convective core. For $Y_0 = 0.4$ the
convective core already develops for $M = 0.9 \mathrm{M_\odot}$.

The mass loss prescription (\ref{eqn:scmassloss}) predicts a higher mass
loss rate on the RGB for hotter stars, which means that the mass loss rate
of helium enhanced stars is higher than that
of helium normal stars.
Nevertheless, the integrated mass loss on
the RGB is lower for helium enhanced stars. This is due to the lower
luminosity at the tip of the RGB, which is in turn due to the helium flash
occurring at lower core mass. This reduces the integrated mass loss mainly
because the mass loss rate is highest near the tip of the RGB and also
because it shortens the RGB lifetime.

The lower core mass for helium ignition is due to an increased temperature
at the base of the hydrogen burning shell.
The easiest way to see why a higher helium abundance in the envelope leads
to an increase of temperature in the core is by shell source homology
\citep{RefsdalWeigert1970,KippenhahnWeigert1990}. The temperature at the
bottom of the hydrogen burning shell scales as
\beq
\frac{T_\mathrm{shell}}{\mu} \propto \frac{M_\mathrm{c}}{R_\mathrm{c}},
\eeq
where the molecular weight $\mu$ is a suitable ``typical'' value in the
burning shell.
Because the core is degenerate there is a relation between $M_\mathrm{c}$
and $R_\mathrm{c}$, say $R_\mathrm{c} \propto M_\mathrm{c}^{-\alpha}$ with
$\alpha > 0$ ($\alpha = 1/3$ for cold degenerate matter). Using primes to
indicate a helium enhanced model, we have
\beq
\frac{T_\mathrm{shell}}{T'_\mathrm{shell}} = \frac{\mu}{\mu'}
\left(\frac{M_\mathrm{c}}{M'_\mathrm{c}}\right)^{1+\alpha},
\eeq
so for the same core mass the helium enhanced model will have a higher
temperature at the bottom of the burning shell ($\mu' > \mu$), which
implies a higher temperature in the core.
An alternative approach that leads to the same conclusion is to apply the
virial theorem for the core and consider the effect of changing the
composition of the envelope
\citep[\emph{c.f.}][$\S$ 30.5]{KippenhahnWeigert1990}.

The lower mass loss on the RGB increases the mass on the horizontal branch
for helium-rich stars of a given initial mass, but because of the
difference in lifetime a coeval helium-rich horizontal branch star will
actually be less massive than a helium normal star.

The core mass at the beginning of the tp-AGB is higher for helium enhanced
stars of the same mass but also for helium enhanced stars of the
same lifetime.
Because of this we expect the white dwarf remnants of helium enhanced stars to
be more massive than the remnants of helium normal stars.
This is an interesting result but the implications of this for the white
dwarf population in globular clusters are beyond the scope of this work.

\section{Structure of the collision products}
\subsection{Lifetime and luminosity and colour functions}
The structure and evolution of the collision products agrees very well with
previous findings \citep{Sillsetal1997,GlebbeekPols2008}, bearing in mind the differences
between helium enhanced and helium normal stars discussed in the previous
section.

\begin{figure}
\begin{center}
\includegraphics[width=0.5\textwidth]{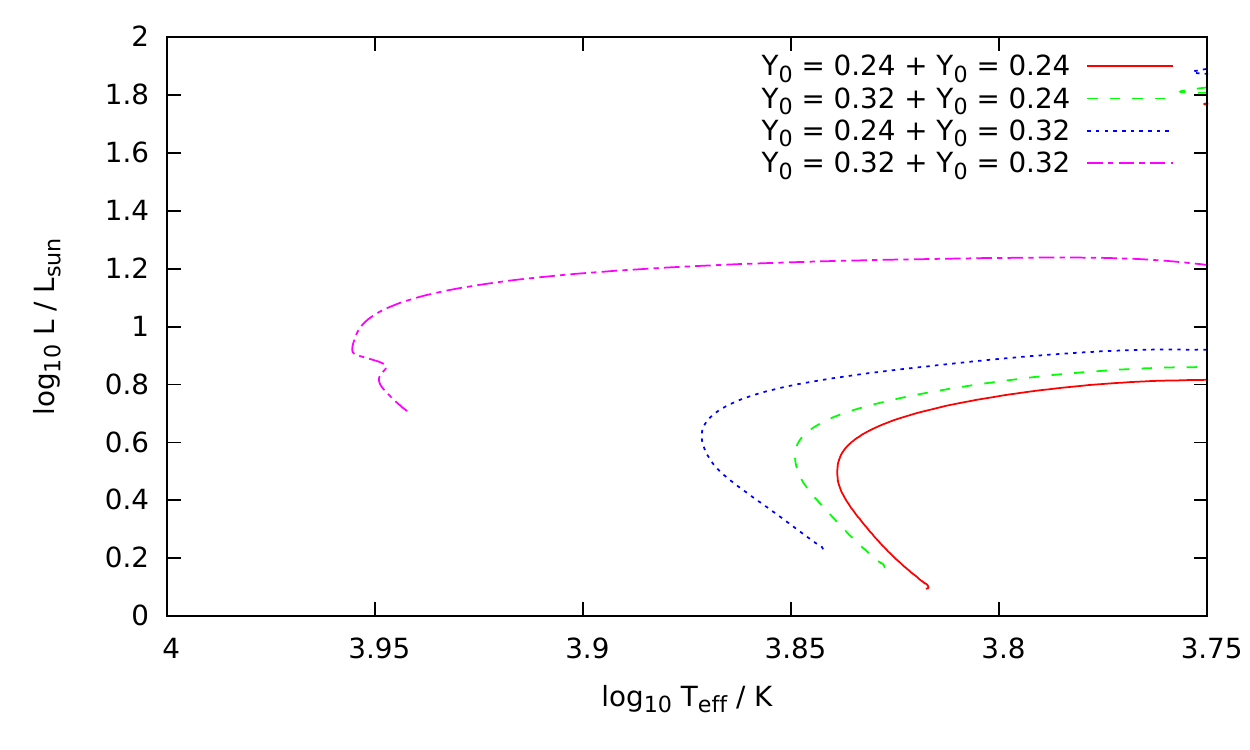}
\end{center}
\caption{Evolution tracks for collisions between an $M=0.6\mathrm{M}_\odot$
and an $M=0.4\mathrm{M}_\odot$ star at 6000 Myr. The initial contraction
phase is not shown for clarity.}
\label{fig:hrd_1msun}
\end{figure}

As can be expected from the ``entropy sorting'' principle, the most
helium-rich object will settle at the core of the collision product.
Figure \ref{fig:hrd_1msun} shows the difference in the evolution tracks
for an $M=0.6 + 0.4\mathrm{M}_\odot$ collision product with different
helium abundances in the primaries: as long as at least one of the stars
has $Y_0 = 0.24$, the evolution tracks look qualitatively the same, but
brighter and bluer with increasing helium abundance in the interior.

The evolution track of a collision product resulting
from $Y_0 = 0.24$ and $Y_0 = 0.32$ progenitors is intermediate between the
evolution tracks of stars with the same mass but an initially uniform
helium composition of $Y_0 = 0.24$ or $Y_0 = 0.32$. 
For parent stars of equal mass the evolution track is similar to that of a
$Y_0 = 0.28$ star of the same mass although it is not as blue.
The evolution of the core, and therefore the lifetime, more closely
resembles that of a $Y_0 = 0.32$ star than that of a $Y_0 = 0.28$ star.

In previous work \citep{Sillsetal1997,GlebbeekPols2008} we showed that the core
of the collision product is formed by the secondary, unless the primary has
evolved sufficiently to reduce its central entropy below that of the
secondary.
This is slightly more complicated in the case of merging stars with
different helium abundances. An unevolved star with a higher helium
abundance will have a lower central entropy than an unevolved star of the
same mass with a normal helium abundance. It will therefore tend to settle
in the core of the collision product, which will resemble a more evolved
star, or the collision product of more evolved parents.
For instance, the result of an $M_1 = 0.6 \mathrm{M}_\odot, Y_0 = 0.24$ and
an $M_2 = 0.6 \mathrm{M}_\odot, Y_0 = 0.32$ collision at $6\,000$ Myr
resembles the collision product of $M_1 = 0.4 \mathrm{M}_\odot, Y_0 = 0.24$
and $M_1 = 0.8 \mathrm{M}_\odot, Y_0 = 0.24$ at $8\,000$ Myr.

In general, it is hard to predict which star will form the core of the
collision product without looking at the structure of the two stars. It is
still true that if the primary is sufficiently evolved, it will provide the
core of the collision product even if the secondary started out with a
higher helium content. However, it will need to be more evolved than in the
situation where the secondary has the same helium content initially.

In either case, the lifetime of the collision product is reduced because
its central helium abundance is higher than the situation where both stars
had a normal helium abundance. This effect is exacerbated by the increase
in the stars' luminosity due to the higher helium abundance in the envelope.

\subsection{Comparison with observed blue straggler populations}
\label{sec:comparison}

We compared the results of our models with the blue straggler
populations in four galactic globular clusters: NGC 2808,
NGC 1851, 5634 and NGC 6093 (M80), see Table \ref{tab:clusters}.
The latter cluster was also studied by \citet{Ferraro2003}.
Observations were taken from the HST WFPC2 database by \citet{Piotto2002}
and blue stragglers were selected using the procedure of
\citet{Leigh2007}.\footnote{Blue straggler numbers for the cores
are given in \citet{Leigh2008}. We use the same procedure to find blue
stragglers in the outer regions of the clusters.  We find 33, 50, 11 and 17
blue stragglers for NGC 1851, 2808, 5634 and 6093 respectively.}
\begin{table}
\caption{Photometric parameters of our comparison clusters, taken from
\citet{Harris1996}.}
\begin{center}
\begin{tabular}{lllll}
\hline
ID & Name & $E(B-V)$ & $(m - M)$ & [Fe/H]\\
\hline
\hline
NGC 1851 &     & 0.02 & 15.47 & $-1.22$\\
NGC 2808 &     & 0.23 & 15.56 & $-1.15$\\
NGC 5634 &     & 0.05 & 17.16 & $-1.88$\\
NGC 6093 & M80 & 0.18 & 15.56 & $-1.75$\\
\hline
\end{tabular}
\end{center}
\label{tab:clusters}
\end{table}

\begin{figure}
\begin{center}
\includegraphics[width=0.5\textwidth]{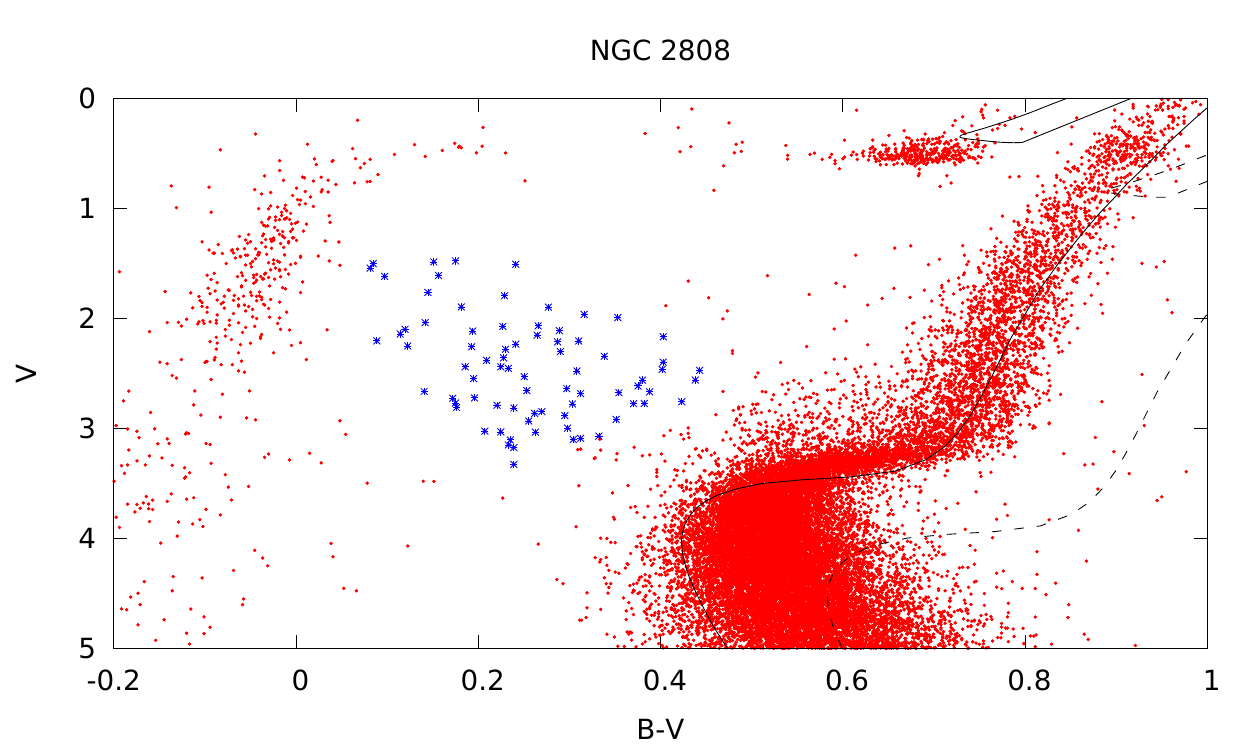}
\end{center}
\caption{Colour-magnitide diagram for NGC 2808, showing the blue stragglers
and a 12 Gyr isochrone for our adopted $E(B-V)$ and $m-M$ (solid line) as
well as the $E(B-v) = 0.23$ (dashed line).}
\label{fig:cmd_ngc2808}
\end{figure}

These clusters were chosen because they all show different indications for
multiple populations.
In the case of NGC 2808, there is an observed splitting of the main
sequence in at least three sequences \citep{Piotto2007}, which is
attributed to the cluster population consisting of distinct sub-populations
with a different helium abundance. It has become common to refer to these
as ``first generation'' and ``second generation'' (or ``third
generation''), where the ``first generation'' refers to stars with a normal
(low) helium abundance and ``second generation'' refers to the helium
enriched population that (supposedly) formed later than the ``first
generation'' out of polluted material.

For NGC 1851, the sub-giant branch is known to be split in at least two
distinct populations \citep{Saviane1998,Milone2008}, which can be explained
if there are two populations with different total CNO abundances
\citep{Cassisi2008,Ventura2009} but with the same helium abundance.
Recently, \citet{Hanetal2009} have presented evidence for a split giant
branch in $U - I$ colours. Their interpretation of this is that there
\emph{is} a Helium enhanced population in NGC 1851 (with $Y = 0.28$) that
also has a slightly higher metallicity and they show that this would only
show up as a split horizontal branch in $V - I$ colours.

Apart from the usual globular cluster abundance anomalies there is no
photometric indication for a ``second generation'' in NGC 6093 or
NGC 5634.

The scenario where some blue stragglers are formed through collisions
between stars of different helium content should therefore apply only to
the case of NGC 2808.

Comparing the photometry from
\citet{Piotto2002} to that of \citet{Walker1999}, there appears to be a
systematic colour shift $\Delta (B-V) \approx 0.1$ to the blue in the
\citet{Piotto2002} data (before applying reddening corrections).
The origin of this offset is unclear, but unless
we correct for it we are unable to properly fit an isochrone to the data.
In practice, we adopted the \citet{Piotto2002} data as it is available and
determined an effective value of $E(B-V) = 0.07$ for our isochrone. We use
this value when comparing our models to the Piotto data.

The reddening for NGC 2808 is somewhat uncertain.
\citet{Walker1999} finds an overall value $E(B-V) = 0.20 \pm 0.02$, but notes
that \citet{Schlegel1998} extinction maps indicate differential reddening
across the cluster. Given the offset between the \citet{Piotto2002} and the
\citet{Walker1999} photometry and the error bars on the reddening, our
effective value $E(B-V) = 0.07$ for the Piotto data appears to be
reasonable.

In order to compare our models with observations we calculate the colour
and luminosity functions predicted by our models at an age of $12
\mathrm{Gyr}$.
The mass spacing of our model grid is not very fine and this results in 
very clear discrete jumps and gaps in the theoretical colour and luminosity
distributions.
To produce smoother distributions, we computed evolution sequences for
collisions between stars with $31$ distinct masses between $0.4$ and $0.8$
(inclusive). For the masses that fall between those in our grid we
interpolate between the neighbouring evolution tracks in a similar way to
\citet{Pols1998}.

In our model sets collisions only happen at discrete time
intervals. In reality, collisions may happen at any time and for our
simulated population we should generate collision models at intermediate
times. We account for this by selecting
for each evolution sequence the portion of the evolution track
that falls within an age range of $12\,\mathrm{Gyr} \pm 500\,\mathrm{Myr}$.
Every point along the evolution track is then assigned a weight
\beq\label{eqn:weight}
W_i = \Delta t_i \phi(m_1) \phi(m_2) \Psi(m_1, m_2, t_\mathrm{coll}),
\eeq
where $\Delta t_i$ is the amount of time spent in the
vicinity of the current point,
$\phi(m) \propto m^{-\alpha}$ is the initial mass function of
\citet{Kroupa2001} and $\Psi(m_1, m_2, t_\mathrm{coll})$ is the collision
probability for star 1 and star 2 at the time of collision
$t_\mathrm{coll}$, which we take to be constant for simplicity.
For model sets D and E the relative abundance of the different populations
is important. We have taken all populations to be equally abundant, so that
for model set D mixed $Y_0=0.24$ and $Y_0=0.32$ collisions are as likely as
collisions between stars with the same helium content.

\begin{figure}
\includegraphics[width=0.5\textwidth]{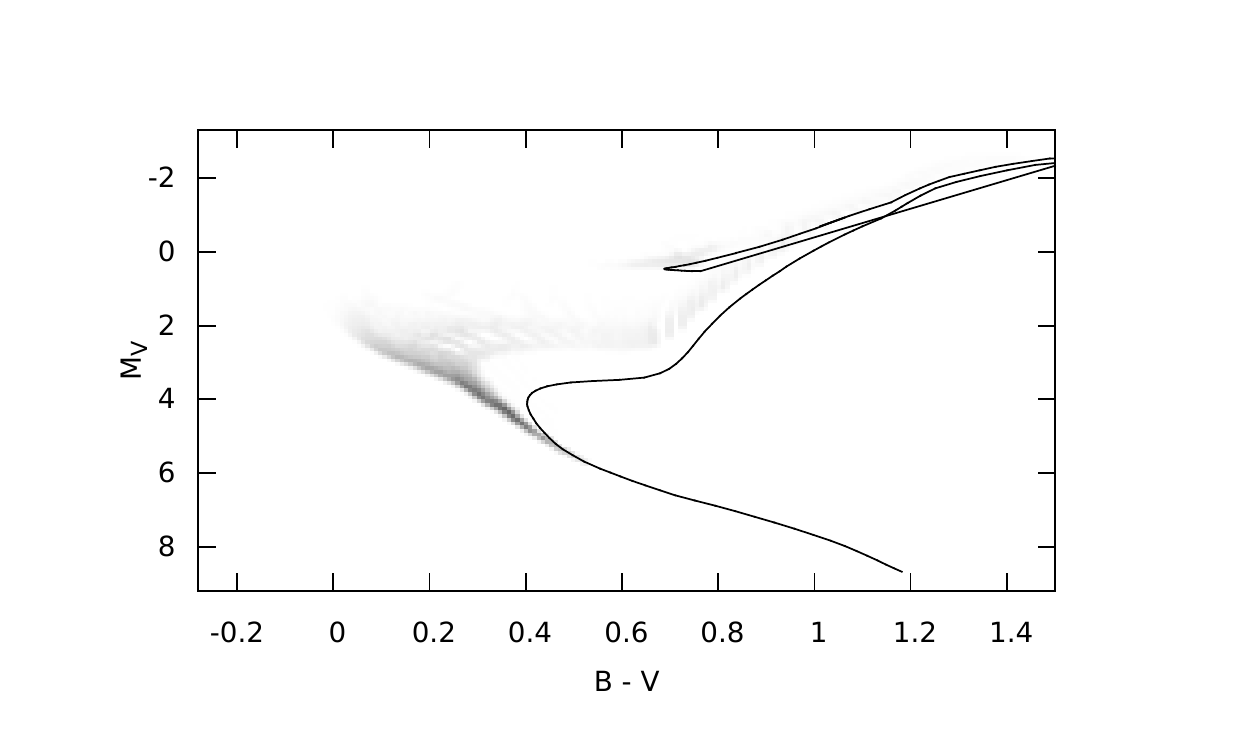}
\caption{Theoretical colour-magnitude diagram for $Y_0 = 0.24$ and
$Z=0.001$ at 12 $\mathrm{Gyr}$. The solid line is a single star isochrone.}
\label{fig:theo_cmd}
\end{figure}

The evolution tracks are then binned in the $B - V$ vs. $M_\mathrm{V}$
plane with the value of each bin being the sum of the weights $W_i$ of
evolution tracks that pass through it. The width of our bins is $0.1$ in
$M_\mathrm{V}$ and $0.089$ in $B-V$. To convert between theoretical $\log
g$, $\log L$ and $T_\mathrm{eff}$ to observational $M_\mathrm{V}$ and $B-V$
we made use of the spectral library by \citet{Lejeune1997, Lejeune1998}.
The resulting colour-magnitude diagram for $Y_0 = 0.24$ is shown in Figure
\ref{fig:theo_cmd} along with a $Y_0 = 0.24$ isochrone.
There are some gaps in the shading in the Hertzsprung gap
between the main sequence and the giant branch where our binning method
misses the stars as the evolve quickly through this region. Because the
time spent in this region of the colour-magnitude diagram is small anyway,
the weight $W_i$ at these points is also small and the presence of these
gaps does not affect the results of our comparison.

The collision products detach from the zero-age main sequence around $B - V
= 0.5$ and fill up the blue straggler region above and to the blue of the
main sequence turn-off. The collision product giant branch appears just to
the blue of the normal single star giant branch and the horizontal branch
is slightly bluer and brighter than the normal single star horizontal
branch.

Observationally, blue stragglers are selected by drawing a selection box in
the colour magnitude diagram, with cut-offs at the low luminosity end to
separate the blue stragglers from the main sequence and at the high
luminosity end to separate the blue stragglers from the horizontal branch.
See \citet{Leigh2007} for a description of an algorithm to define these
selection boxes consistently between different clusters. In some
cases (most notably NGC 6093 and NGC 1851) the resulting selection box
includes one or two blue stragglers that are substantially ($>0.1
\mathrm{dex}$) fainter than the rest. This strongly affects the shape of
the luminosity function at the low-luminosity end and we have
redrawn our selection box to reject these stars from both the observations
and our models.

To compare our models to the observations, we have selected the blue
stragglers from our models using the same selection box as was used to
select the observed blue stragglers.
These selection boxes vary slightly from cluster to cluster, so that
the range of colours and luminosities spanned by the blue stragglers also differs slightly from cluster to cluster.

\begin{figure}
\includegraphics[width=0.5\textwidth]{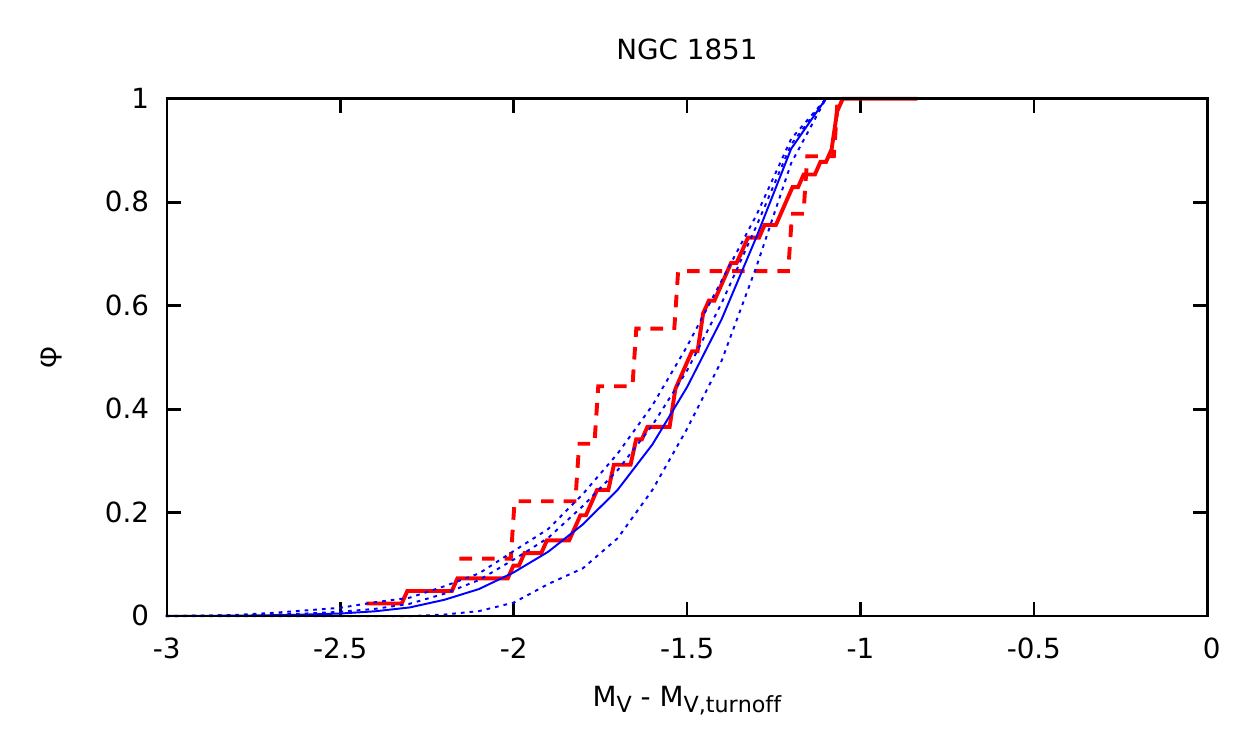}\\
\includegraphics[width=0.5\textwidth]{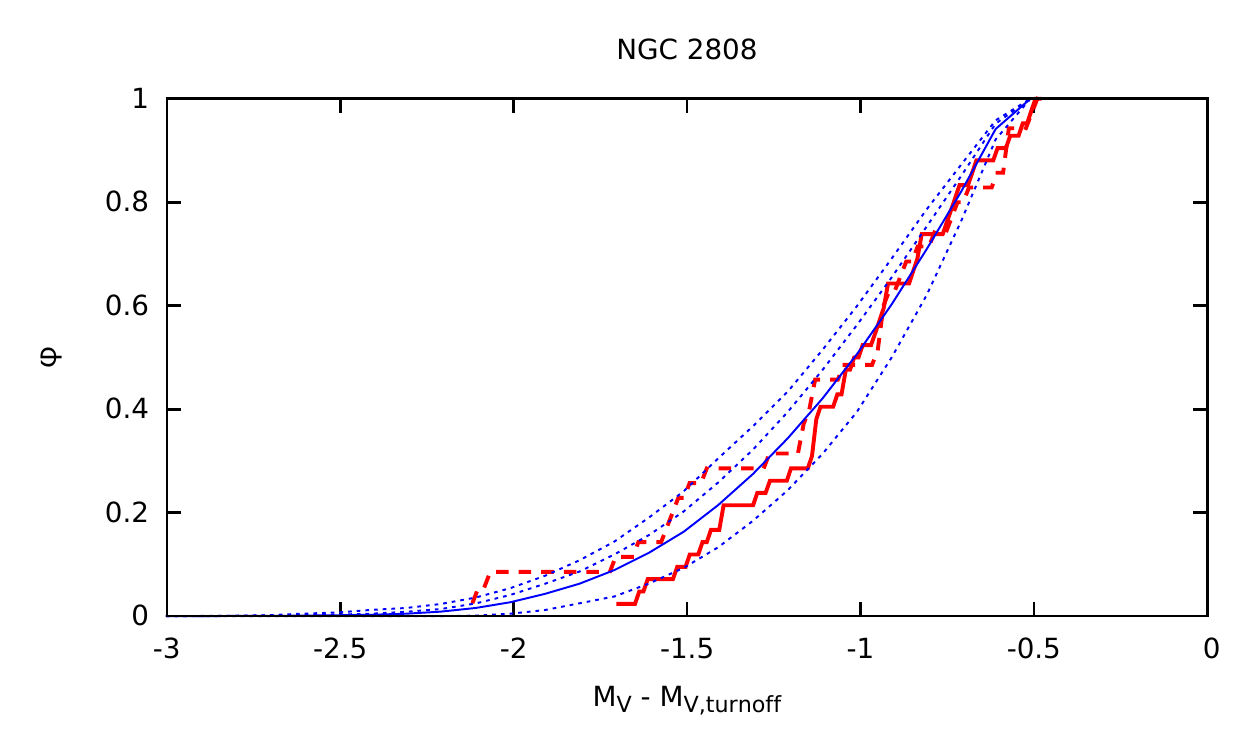}\\
\includegraphics[width=0.5\textwidth]{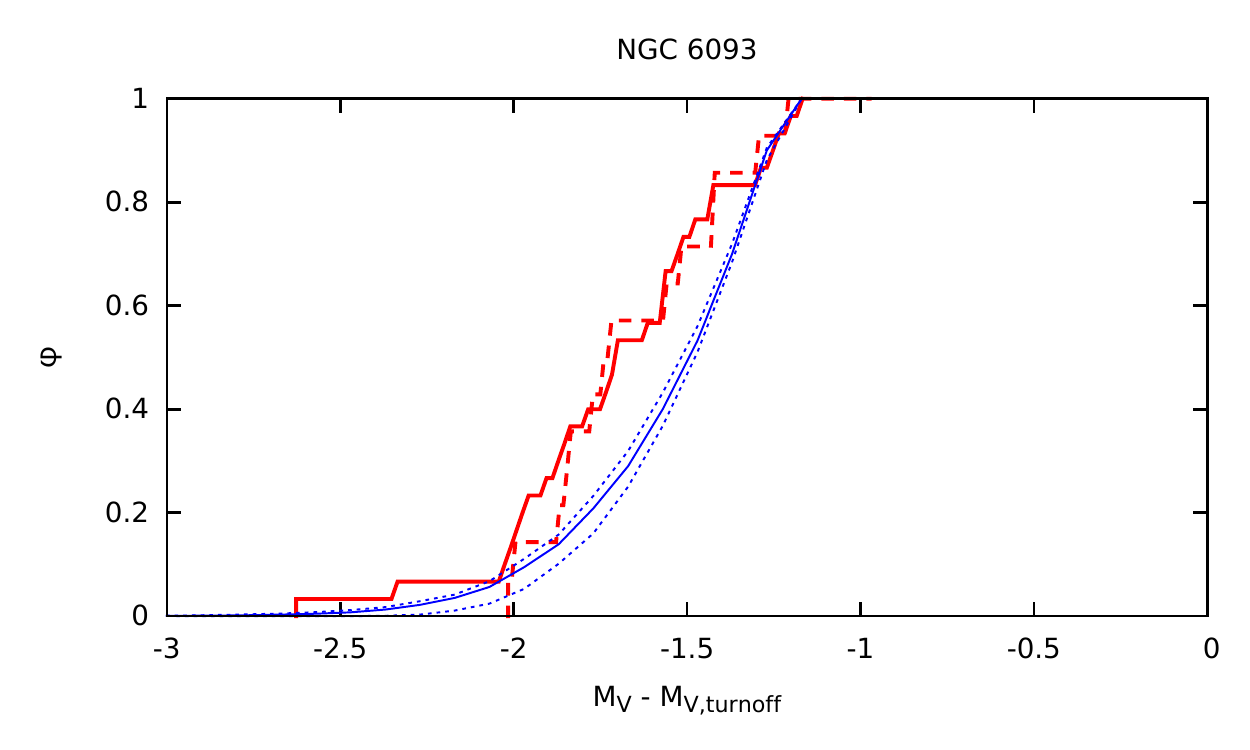}\\
\includegraphics[width=0.5\textwidth]{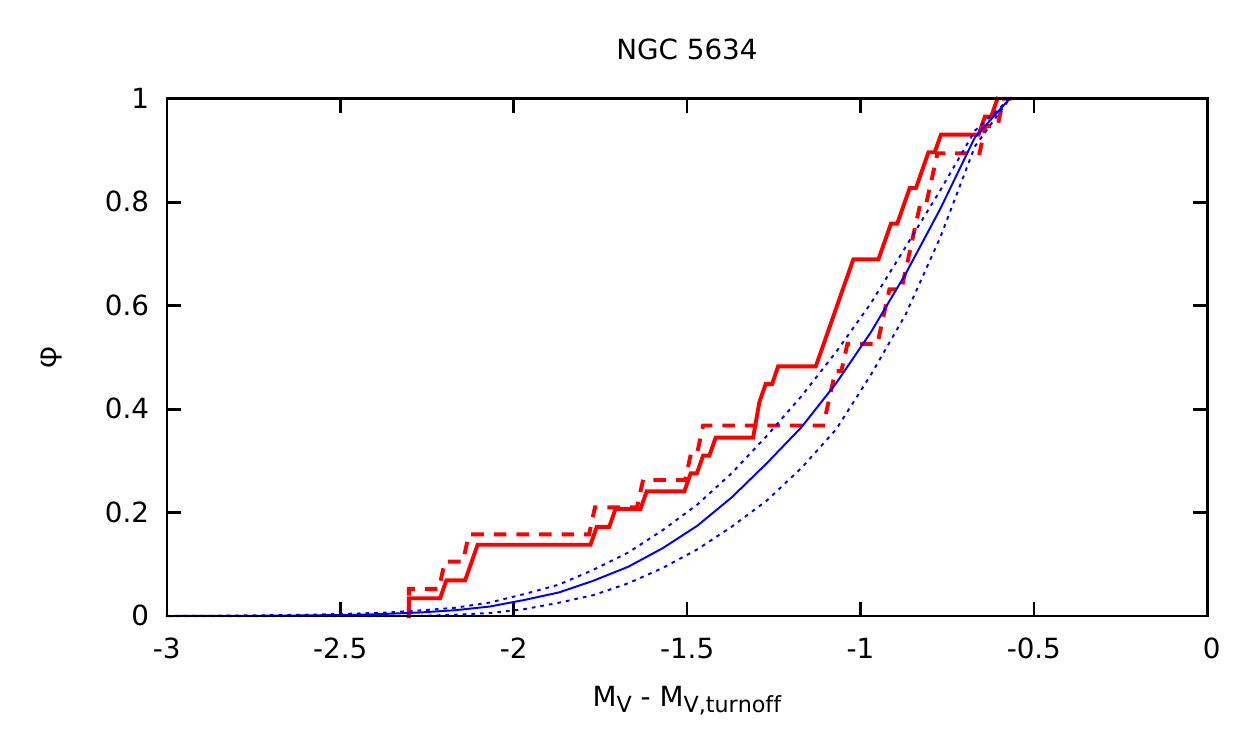}
\caption{Cumulative luminosity functions $\varphi$ for the observed
blue straggler populations in the whole cluster (thick red solid
line), in the cluster core (thick red dashed line) and the luminosity
function for the collision models (thin blue lines).
The solid line is for model set D, the dotted lines are for $Y_0 = 0.40$
(model set C), $Y_0 = 0.24$ (model set A) and $Y_0 = 0.32$ (model set B)
respectively from left to right.}
\label{fig:luminosity_functions}
\end{figure}

\begin{figure}
\includegraphics[width=0.5\textwidth]{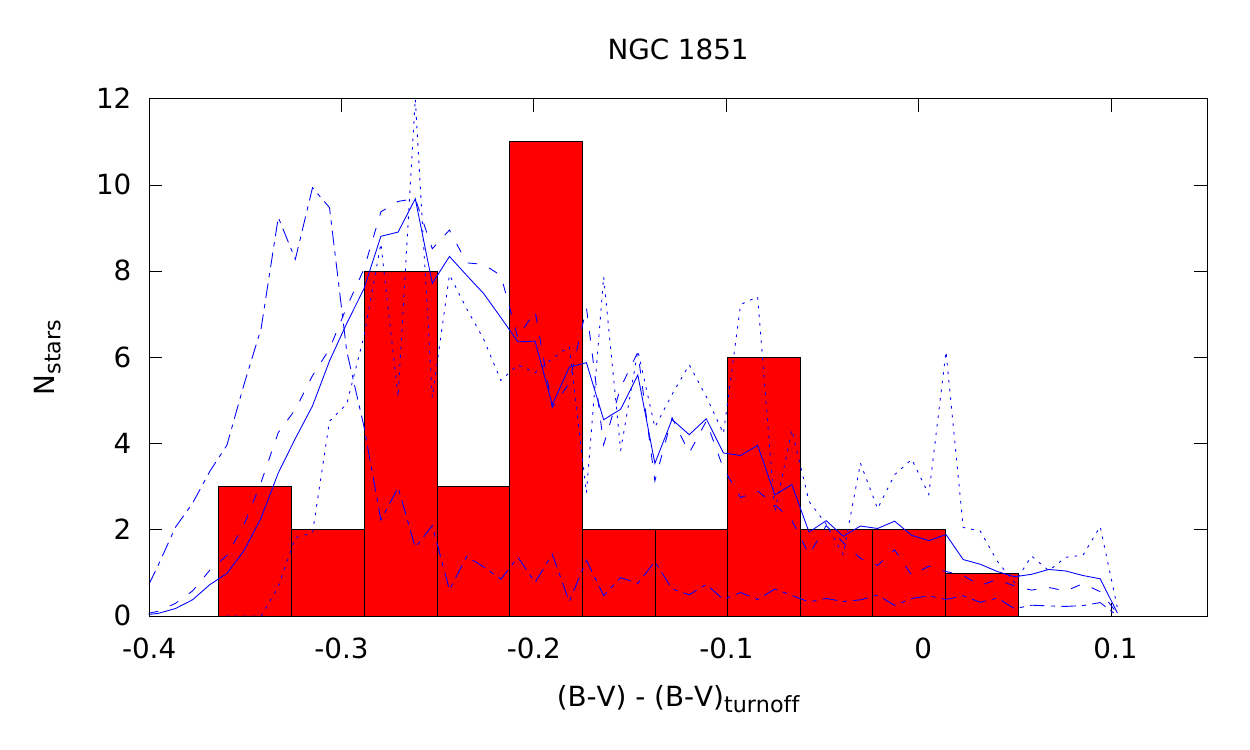}\\
\includegraphics[width=0.5\textwidth]{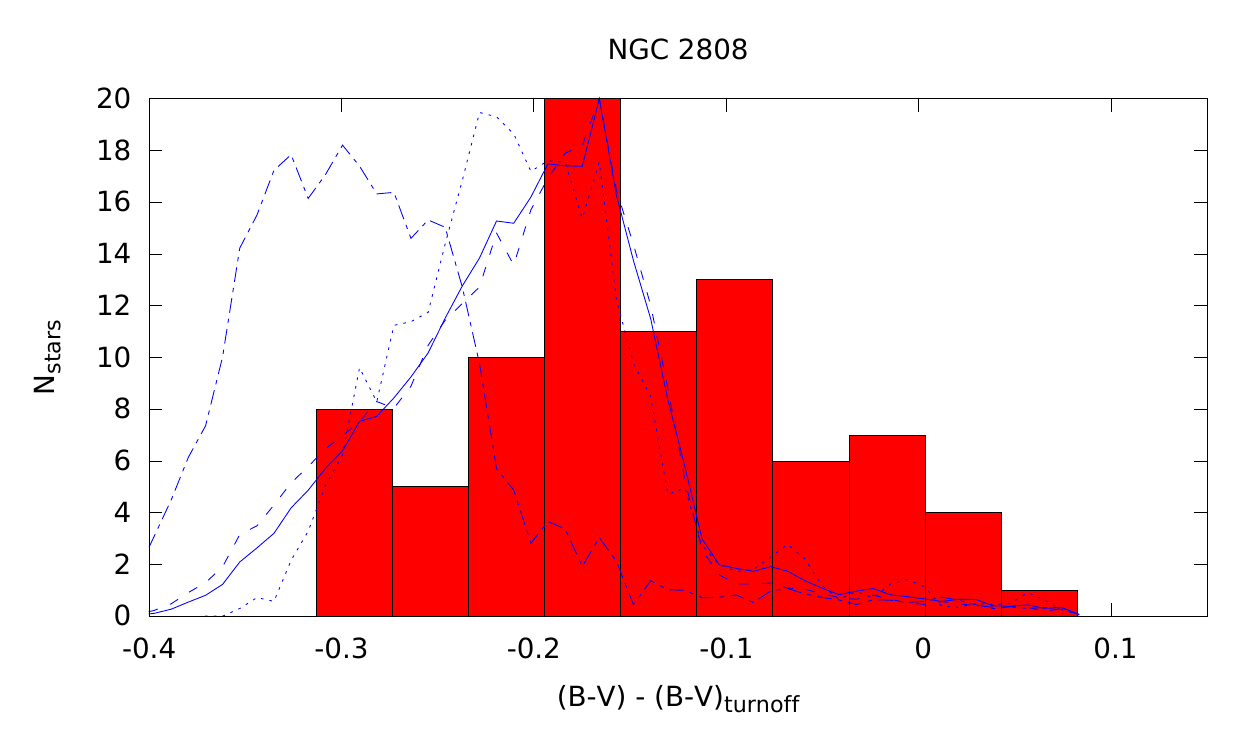}\\
\includegraphics[width=0.5\textwidth]{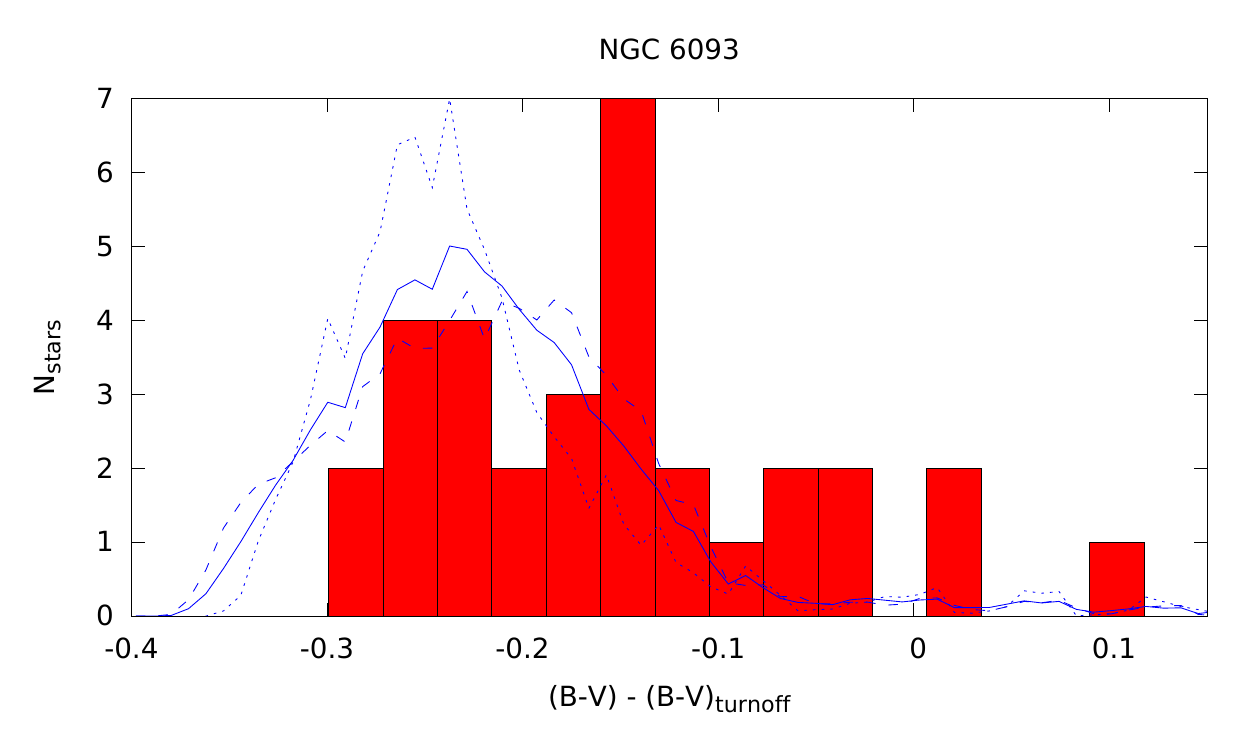}\\
\includegraphics[width=0.5\textwidth]{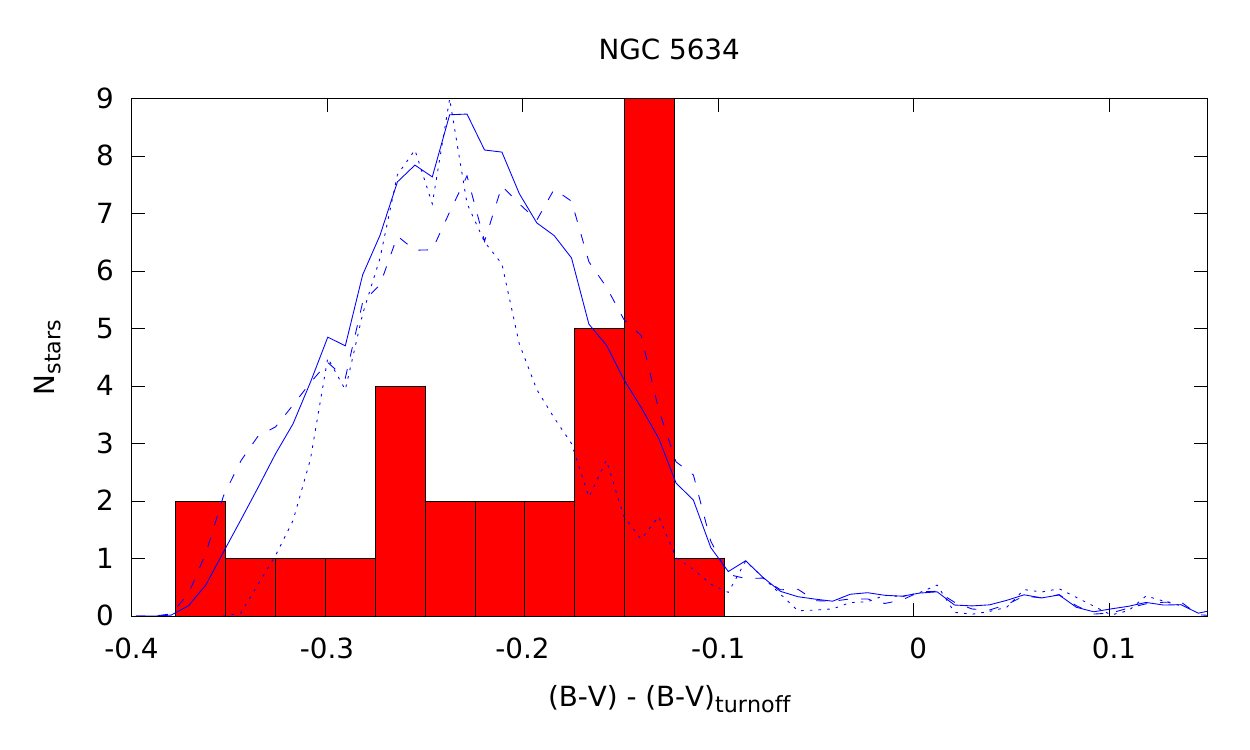}
\caption{Total $B-V$ colour distributions of the observed blue straggler
populations (histogram) and the colour distributions for the collision
models (curves).  The solid line represents model set D. The dashed, dotted
and dash-dotted lines represent model sets A -- C respectively.}
\label{fig:colour_functions}
\end{figure}

We compare our $Z = 0.001$ ($[\mathrm{Fe}/\mathrm{H}] = -1.3$) models with
NGC 1851 and NGC 2808, which have a slightly higher metallicity (Table
\ref{tab:clusters}).
Our $Z = 0.0003$ ($[\mathrm{Fe}/\mathrm{H}] = -1.82$) models are compared
with NGC 5634 and NGC 6093, which straddle this metallicity.

The (observed and model) blue straggler luminosity functions are shown in
Figure \ref{fig:luminosity_functions} and the colour distributions are
shown in Figure \ref{fig:colour_functions}.
Table \ref{tab:goodness_of_fit} gives the probability that the observed
luminosity function is drawn from model sets A, B, C, D or E (defined
in Table \ref{tab:grid}), as determined by a standard K-S test.
The listed value $p_0$ is the probability using the observed value for
$M-m$. The value $p'$ was derived by treating the distance modulus as a
free parameter that was then derived from fitting to the data. The
difference between the observed value of $M-m$ (listed in Table
\ref{tab:clusters}) and the best fitting value for $M-m$ is listed as
$\Delta(M-m)$.
We will discuss each of the clusters in turn.

\begin{table}
\caption{K-S test probabilities comparing the theoretical and
observed blue straggler luminosity functions for the entire
cluster.}
\begin{center}
\begin{tabular}{lllll}
\hline
Cluster & Model set & $p_0$ & $p'$ & $\Delta (M-m)$ \\
\hline
\hline
NGC 1851 & A & 0.45 & 0.75 & $+0.03$ \\
         & B & 0.13 & 0.30 & $-0.02$ \\
         & C & 0.43 & 0.90 & $+0.05$ \\
         & D & 0.45 & 0.61 & $+0.01$ \\
         & E & 0.43 & 0.66 & $+0.02$ \\
\\
NGC 2808 & A & 0.15 & 0.82 & $+0.09$ \\
         & B & 0.19 & 0.45 & $-0.09$ \\
         & C & 0.03 & 0.73 & $+0.14$ \\
         & D & 0.62 & 0.89 & $+0.03$ \\
         & E & 0.45 & 0.88 & $+0.05$ \\
\\
NGC 5634 & A & 0.35 & 0.75 & $-0.12$ \\
         & B & $< 10^{-3}$ & 0.52 & $-0.20$ \\
         & D & 0.05 & 0.69 & $-0.17$ \\
\\
NGC 6093 & A & 0.12 & 0.40 & $-0.04$ \\
         & B & 0.01 & 0.18 & $-0.07$ \\
         & D & 0.04 & 0.27 & $-0.06$ \\
\hline
\end{tabular}
\end{center}
\label{tab:goodness_of_fit}
\end{table}

\subsubsection{NGC 1851}
For NGC 1851 most of our model luminosity functions fit the observations
about equally well ($p\sim 0.45$) with the exception of the helium enhanced
$Y_0 = 0.32$ models (model set B). If we allow for a shift in $M-m$ then
the best fits are for model set C ($p=0.90$) and A ($p=0.75$).

The high luminosity end of the distribution best fits the pure $Y_0 = 0.24$
models (model set A), with $p>0.999$ if we only select blue stragglers
brighter than $M_\mathrm{V} = 2.7$.
We can match the fainter end of the luminosity function, which then matches
best with the $Y_0 = 0.4$ collision models (model set C) with $p = 0.81$.
However, in that case our models do not match the high luminosity end of the
luminosity function.
The luminosity function for the core is clearly brighter than for the
cluster as a whole, which could be a signature of mass segregation.
However, the number of stars is small so it is hard to draw any
firm conclusions.

The predicted colour distribution from our models gives a tolerable fit to
the observations. We predict more stars between $B-V = 0.2$ and $0.3$, but
the number of observed stars in this colour range is small. The
predicted location of the peak in the colour distribution may be somewhat
too blue.
A K-S test for the cumulative colour distribution functions is not very
conclusive. Model set C is too blue on average and has the
lowest probability, of $0.52$. The next lowest probability is $0.97$ for
model set E. All other model sets have probabilities $>0.99$.

Taking into account both colour and luminosity, model set A
($Y_0 = 0.24$) is most consistent with observations for NGC 1851.

\subsubsection{NGC 2808}
For NGC 2808 the best agreement ($p = 0.62$) is obtained for a mixed
population of collision products involving $Y_0 = 0.24$ and $Y_0 = 0.32$
stars (model set D), although good agreement with the pure $Y_0 = 0.24$ and
a mixture of all collision products can also be obtained by allowing for a
shift in the distance modulus. The combination of $Y_0 = 0.24$ and $Y_0 =
0.32$ marginally remains in best agreement with the observations, however.
Interestingly enough, the high-luminosity end of the luminosity function is
in better agreement with the pure $Y_0 = 0.32$ collisions, which is in line
with expectations.

The core luminosity function seems to fit slightly better with a pure $Y_0
= 0.24$ component for luminosities that are one magnitude
brighter than the turnoff, which is compatible with mass segregation since
stars of $Y_0 = 0.24$ are expected to be more massive than stars with $Y_0
= 0.32$ on average.

The range of colours spanned by our collision models agrees well with the
observations, and both place the peak between $\Delta(B - V) = -0.2$ and
$-0.1$. The observations may fall off more quickly on the blue side, but
the difference
is within the expected error based on the number of observed stars in the
colour bins. On the other hand, the observations clearly show an excess of
blue stragglers to the red of $\Delta(B - V) = -0.1$. In our models the
stars in this colour range are post-main sequence objects that are in the
Hertzsprung gap. Similar discrepancies between models and observations have
been noted before \citep[\emph{e.g.}][]{Sills2000}.
Extra mixing, for instance due to rapid rotation, offers one possible way
to extend the lifetime of stars in this region. Convective overshooting, as
noted before, is ineffective in removing this discrepancy.
Another possibility is that at least some of the stars in this region are
unresolved binaries.

A K-S test of the cumulative colour distribution gives about equal
probability of $0.70$ to both model sets B and D, followed by model set A
with a probability of $0.62$. The pure $Y_0 = 0.4$ models (model set C) has
a probability $< 0.001$. Combining both the colour and luminosity
information, our model set D best describes the blue stragglers in NGC
2808.

\subsubsection{NGC 6093}
None of our luminosity functions fit particularly well.
The best fitting model is model set A, although model set D is not much
worse. By varying the distance modulus, model set A can be made to fit a
bit better, while the fit for model set D does not improve significantly.

The colour distributions for the different model sets are all
very similar. The observations show a peak that is $0.15$ magnitudes
bluer than the turnoff. As with NGC 2808, there are more observed blue
stragglers redward of $\Delta(B-V) = -0.1$ than predicted by our models. By
contrast, no blue stragglers are observed blueward of $\Delta(B-V) = -0.3$.
The $Y_0 = 0.24$ models fit slightly better ($p = 0.39$) than the mixed
$Y_0=0.24; 0.32$ models ($p = 0.25$) but the picture is not very clear.

\subsubsection{NGC 5634}
Again, none of our models fit very well.
Model set A ($Y_0 = 0.24$) gives the best agreement with the
observations, although model set D is also not a bad fit if we vary the
distance modulus. The theoretical
luminosity function falls off perhaps a little too quickly for higher
luminosities.
The colour distributions are not very different for the three model sets
and all match about equally well. The observations show a peak 0.1 dex
bluer than the turnoff that is not present in the models.

\subsection{Late evolutionary phases}
The late (post-main sequence) evolution of collision products is
interesting for two reasons. First of all it allows us to test our
understanding of the subsequent evolution of collision products by
comparing the observed distributions and properties of evolved blue
stragglers (post-blue stragglers) with the models. Second of all, it may
allow us to probe the dynamical history of the cluster over a longer time
interval than can be accomplished with the blue stragglers alone.

The most promising post-main sequence evolutionary phase to identify
stellar collision products is during core helium burning. Observationally,
the collision products are then expected to lie above the horizontal branch
\citep{Renzini1988,FusiPecci1992,Ferraro1999,Sillsetal2009}.

We can derive a selection box for evolved collision products by using the
evolution models.
The first step is to determine the minimum and maximum values of
$M_V$ and $B - V$ during core helium burning for all of our collision
models. These define a selection box in the colour-magnitude diagram that
contains the core helium burning phase of all our collision products.
However, for some models this selection box will now encompass more than just
the core helium burning phase. Due to the difference in helium content, the
colour of the giant branch is different for different models in our set.
We therefore restrict our selection box so that it does not overlap with
the red giant branch in any of our models. We also impose a cut to remove
the cluster horizontal branch. This way, we aim to select only stars that
are unambiguously evolved blue stragglers.
Because we have restricted the selection box, we will not capture all of
the core helium burning phase for all collision products, and we may miss
$\sim 20\%$ of the core He burning collision products by using this
narrower selection box.
In the $\log_{10} T_\mathrm{eff}$ / $\log_{10} L / L_\odot$ plane our
selection box is very similar to that of \citet{Sillsetal2009}.

We can define a selection box for the AGB in the same way as for the core
helium burning phase, but this selection box turns out to be very narrow.
Since the expected ratio of blue stragglers to AGB post-blue stragglers is
also very high, $\sim 1000$, it is virtually impossible to compare with
observations. Therefore, we only compare the ratio of blue stragglers to
core helium burning post-blue stragglers.

The predicted ratio of blue stragglers to evolved blue stragglers is
typically $20$ -- $50$. Observationally, the ratios are about $10$, but
there are only a few stars in the evolved blue straggler selection box
and numbers are very sensitive to whether any particular star is classified
as an evolved blue straggler or not.

A reasonable agreement between models and observations was reached by
\citet{Sillsetal2009} based on the average of the horizontal branch and
main sequence lifetimes of their collision products, as long as they
selected only the brighter (more massive) collision products.
Our approach differs from theirs in two respects: first of all we calculate
a population of collision products where each collision product is assigned
a weight that depends on the IMF probability of the parent stars as opposed
to taking a straight average. Second of all we calculate our population
ratios, for both the models and the observations, based on selection boxes
in the colour magnitude diagram, not on the relative duration of the
evolution phases themselves.

The first of these has the effect of giving more importance to the lower
mass collision products, which will tend to increase the predicted ratios.
The effect of using selection boxes that will only capture part of the
evolution rather than comparing the lifetimes directly will similarly tend
to increase the blue straggler to post-blue straggler ratio.
Neither of these effects should affect the comparison between our
models and the observations, however, because blue stragglers and evolved
blue stragglers are selected consistently between the observations and in
the models -- if the models reflect the observed population, then these
population ratios should come out the same as long as the selection
criteria are the same.
However, the small number of observed stars that actually fall within our
selection box make it hard to draw any firm conclusions from this
comparison.

\section{Summary and Discussion}

We present new evolution calculations for stellar collision products, where
we have allowed the progenitor stars to have a different helium abundance.

Our collision models do a reasonable job of reproducing the observed blue
straggler luminosity function and colour distribution. We predict, perhaps,
too few blue stragglers at the red edge of the observed distribution.
For NGC 1851, the best agreement is obtained for a single population of helium
normal stars ($Y_0 = 0.24$) while for NGC 2808 the best fit is obtained
with a population of mixed $Y_0 = 0.24$ and $Y_0 = 0.32$ stars. These
results are what we would expect in light of observations of
multiple populations in these clusters.
A lower metallicity set of models agrees with NGC 6093 and NGC
5634 in the sense that the best fitting models are those without a helium
enhanced population.
The observed population ratios of blue stragglers to evolved blue stragglers
is larger than is seen in the observations, but the number of observed
evolved blue straggler candidates is small.

Recently, \citet{Hanetal2009} reported the presence of two distinct
populations in NGC 1851, one helium normal and one helium enhanced ($Y =
0.28$). However, they also report that the helium-rich population has a
higher metallicity, which offsets the colour changes due to helium
enhancement, at least in $V - I$ colours.
Since all our models have the same metallicity, there is no corresponding
compensation for the shift in colour due to helium enhancement. To compare
our models more directly we would need to allow for a similar shift in
metallicity.

Although the agreement between our models and the observations is
reasonable, the agreement is not perfect and the models
presented here are not complete.
There are two obvious improvements that can be made.
The first of these has to do with the blue straggler formation rate, which
we have effectively taken to be constant and independent of the helium
abundances of the two colliding stars. By allowing the collision rate
$\Psi$ in (\ref{eqn:weight}) to vary with $t_\mathrm{coll}$ we have more
freedom in shaping the colour and luminosity functions -- in fact, if we
want to use blue stragglers (and possibly evolved blue stragglers) to learn
about the dynamical history of their host cluster, then it is essential
that we allow this factor to vary. Simulations of clusters hosting
multiple populations, like those of \citet{DErcole2008} may serve as a
guide for how this should be done. Ideally, the collision rate should be
directly determined from dynamical simulations of cluster evolution. A software
environment that might be especially suited for this task is the MUSE
software package, which is in active development \citep{muse2009}.

The second improvement that can (and should) be made is that we should
consider binaries in addition to single star models.
For the blue stragglers, this is an obvious addition because binary mass
transfer is an alternative scenario for blue straggler production
\citep[\emph{e.g.}][]{ChenHan2009}.
However, binaries are also important in another respect: observationally,
it is not possible to distinguish the light of the two stars in a binary
system, which may place the system in an unusual point in the colour
magnitude diagram that is hard to reproduce with single star evolution
tracks. This becomes especially important when we look for evolved blue
stragglers because a blend of a horizontal branch star and
a red giant may appear in the same region of the colour-magnitude diagram
as the evolved blue stragglers. 
Such unresolved binaries can be recognised by multi-wavelength
photometry because the colour of the unresolved binary will not change in
the same way as that of a single star, and the binary will move to a
different region of the colour-magnitude diagram: a normal star will stay
close to other stars of a similar spectral type, but an unresolved binary
will move closer to the position of one of its unresolved components.

Evolved blue stragglers offer an interesting possibility to test our
understanding of blue straggler evolution, but because the number of
observed post-blue straggler candidates is small it is especially important
to understand how this region of the colour magnitude diagram may be
influenced by the presence of binaries. 
Both normal binary interaction and collisions can increase the actual number of 
stars in the evolved blue straggler region without increasing the number of
blue stragglers.
Unstable binary mass transfer from a red giant to an unevolved main
sequence star companion can lead to a spiral in and merger of the two stars
if the envelope is not ejected. A collision with a red giant will have a
similar result.
The red giant core is likely to remain intact during the merger so the
merger product will still have a degenerate helium core and evolve like a
more massive red giant. Such a merger would show up as an ``evolved blue
straggler'' despite never having been a blue straggler itself.

Our population models are consistent with observations of multiple
populations in the sense that helium enhanced model sets fit best with
clusters (in particular, NGC 2808) where helium enhancement has been
inferred from the observations.
We hope that with the inclusion of a population of binaries, population
models such as we have presented in this paper could be used not only to
test our understanding of cluster dynamics but also to get a better handle
on the nature of the multiple populations that are now observed in star
clusters.
Future cluster simulations that include an accurate treatment of the
evolution of stellar collision products as well as multiple populations
will be an important diagnostic tool and we plan to make our models
available for such a study.

\begin{acknowledgements}
We thank Peter Anders for his help with the spectral libraries. We thank
the Kavli Institute for Theoretical Physics for their hospitality during
the Evolution of Globular Clusters programme where this work was begun.
Finally we thank the anonymous referee for useful comments that helped to
improve the clairity of this paper.
KITP is supported in part by the National Science Foundation under Grant
No.\ PHY05-51164. AS is supported by NSERC.
\end{acknowledgements}

\bibliography{helium_collisions}

\end{document}